\documentclass{llncs}
\usepackage{makeidx}  

\usepackage{epsfig}
\usepackage{amsmath}%
\usepackage{amssymb}

\usepackage[lined]{algorithm2e}
\begin{document}
\frontmatter          
\pagestyle{headings}  
\addtocmark{Hamiltonian Mechanics} 
\mainmatter              
\title{Combining Stream Mining and Neural Networks for Short Term Delay Prediction}
\titlerunning{Stream Mining and  Neural Networks for Short Term Delay Prediction}  
%
\author{Maciej Grzenda \and Karolina Kwasiborska \and
Tomasz Zaremba}
\authorrunning{Maciej Grzenda et al.} 
%
\tocauthor{Maciej Grzenda, Karolina Kwasiborska,
Tomasz Zaremba}
\institute{Warsaw University of Technology, Faculty of Mathematics and Information Science, 00-662 Warszawa, ul. Koszykowa 75, Poland\\
\email{\{M.Grzenda, K. Kwasiborska, T. Zaremba\}@mini.pw.edu.pl},\\ WWW home page:
\texttt{http://www.mini.pw.edu.pl}
}

\maketitle              

\begin{abstract}
The systems monitoring the location of public transport vehicles rely on wireless transmission. The location readings from GPS-based devices
are received with some latency caused by periodical data transmission and temporal problems preventing data transmission. This negatively affects identification of delayed vehicles.
The primary objective of the work is to propose short term hybrid delay prediction method. The method relies on adaptive selection of Hoeffding trees, being stream classification technique and multilayer perceptrons. In this way, the hybrid method proposed in this study provides anytime predictions and eliminates the need to collect extensive training data before any predictions can be made. Moreover, the use of neural networks increases the accuracy of the predictions compared  with the use of Hoeffding trees only.

\end{abstract}

\section{Introduction}

The idea of Internet of Things relies on connecting objects or things to the internet \cite{tsai2014}. 
Taking into account frequently large number of sensors and data they produce, IoT systems have become one of major sources of high volume, velocity and variety data, frequently referred to as {\em big data} \cite{qin2016}. As a consequence, attempts to use {\em big data} frameworks for  IoT data processing have been observed recently. 
Examples go beyond data storage and include also IoT data stream processing. These include knowledge-based systems, such as the system proposed by S. Kolozali et al. in \cite{kolozali2014}. 
In the case of  big data systems, frequently Lambda architecture proposed by N. Marz and described in \cite{marz2013} is applied for the systems processing IoT data. In line with Lambda design pattern newly arriving data streams are placed in parallel in persistent storage and forwarded to stream processing module, in many cases relying on a stream processing system \cite{hesse2015} also referred to as a stream processing engine (SPE). In the case of data storage, Apache Hadoop is a frequent choice, while stream processing usually relies on one of SPEs such as Apache Spark or Apache Flink.

What is of particular interest for this study is intelligent processing of data streams arriving from IoT networks involving the use of prediction techniques. The case study this work concentrates on is the processing of vehicle location data acquired from hundreds of spatially distributed sensors, and transmitted via wireless links.  Due to limited availability of wireless bandwidth, the readings are observed to be incomplete and delayed. This is not unique for the location data streams.
In their recent survey on data-centric IoT \cite{qin2016}, Y. Qin et al. notice that intrinsic features of IoT data include dynamics caused by changing location of many objects, and uncertainty related to limited precision and completeness of the data combined with its redundancy.

At the same time the processing of continuously arriving data raises a question of which category of machine learning techniques to select. 
As a consequence of a growing emphasis on near-real time stream processing rather than periodical batch analysis, the stream-oriented techniques attract growing attention. In particular, stream classification techniques such as Hoeffding trees \cite{domingos2000} or recently proposed adaptive model rules \cite{almeida2013} are used to develop classification and regression models, which continuously evolve following the changes in data streams.  Some of the key aspects of these techniques are the ability to evolve the model after every instance, the ability to process unbounded volumes of training data available in the form of data streams or anytime property defined as the ability of the model to be ready to be applied at any point
between training examples \cite{bifet2009data}. In addition, some stream mining methods address intrinsic non-stationary phenomenon of many data streams and are able to respond to concept drift \cite{ditzler2015}. 


This study proposes the use of stream classification to predict delays of public transport vehicles. To perform such predictions, location data streams have to be pre-processed, fused and integrated with schedule data first.  
Next, the input data for delay prediction has to be prepared and used to develop and tune stream classification models. The results of model evaluation show the merits of stream mining, but also suggest possible accuracy improvements. These are successfully provided by hybridisation of soft computing approach, namely the use of multilayer perceptrons (MLP) with stream mining, which is proposed in this study.

The remainder of this work is organised as follows. First, data stream processing including data fusion is described in Sect. \ref{sec:fusion}. This includes the use of stream processing frameworks. Next, the problem of short term delay prediction is formulated. In particular, the need for such prediction is explained and possible input data is considered. In Sect. \ref{sec:method}, a proposal of the use of stream mining for delay prediction and a proposal of the novel hybrid technique is made. This is followed by data description and analysis of results made in Sect.~\ref{sec:results}. Finally, the results of this work are summarised in Sect. \ref{sec:conclusions}.

\section{Data fusion with big data frameworks}
\label{sec:fusion}
\subsection{Data acquisition}
This work relies on public transport data made available by the city of Warsaw. The data is provisioned through the Open Data portal of the city available at \url{http://api.um.warszawa.pl} and is publicly available. It takes the form of GPS-based location records produced in near-real time manner for individual trams.
  GPS readouts are inevitably noisy with minor fluctuations of the location observed, even when vehicles remain stationary. Moreover, some location updates arrive with latency to server backend due to variety of problems such as obstacles preventing radio transmission.
Since the location of public transport vehicles constantly changes, these data are served through REST-style API, which makes it possible to poll for most recent vehicle location known to server backend.

\subsection{Data fusion with stream processing framework}
\begin{figure}
\includegraphics[width=\textwidth]{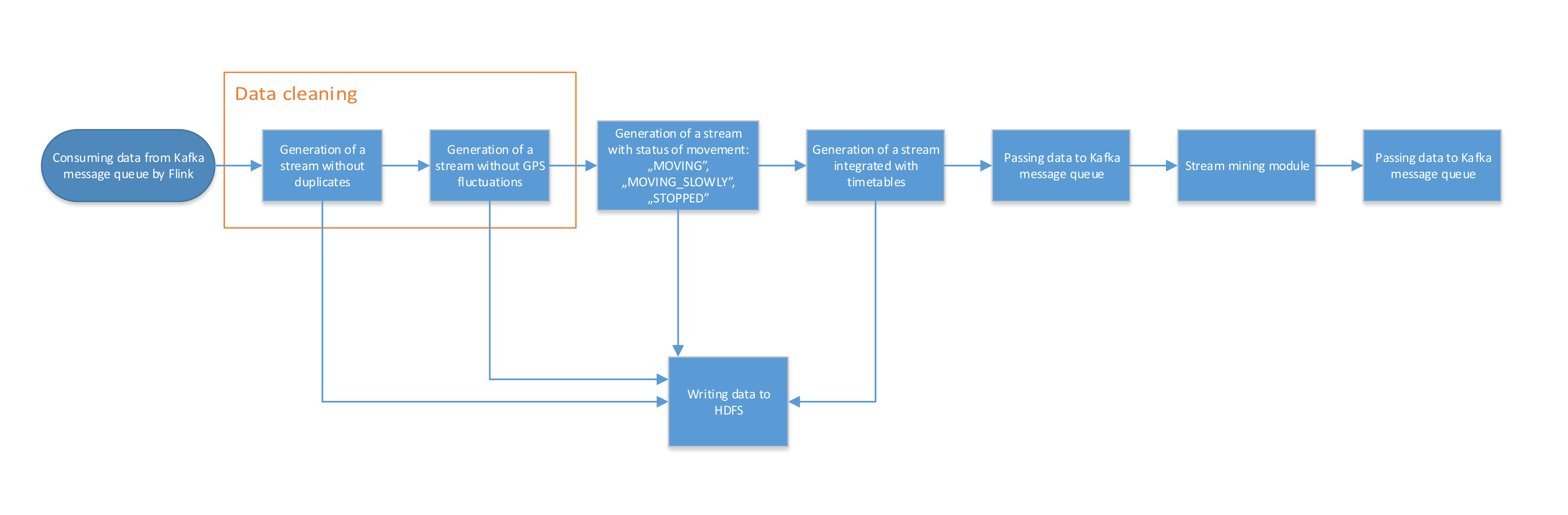}
\caption{The development of location and delay data streams}
\label{fig:streaming}      
\end{figure}

Before stream mining method is described, it is important to document the process of creating the data stream containing the input data used for delay prediction. First of all, most recent location of trams is polled from the aforementioned Warsaw API platform every 30 seconds. These data are forwarded to Apache Kafka streaming platform used to host the queue of location records. These records comprise on a data stream of data tuples processed by a task running under the control of  Apache Flink. The overview of the process of collecting the data from Apache Kafka and performing data pre-processing and data fusion is provided in Fig. \ref{fig:streaming}.

Based on raw location data streams, secondary data streams are constantly produced. First of all $t_\mathrm{R}$
i.e. the system time of receiving the data from API is added to stream instance. Non-negligible difference between this time and the time of capturing location at a vehicle $t_\mathrm{C}$ is observed. Moreover, the same location record can be received many times from Open Data portal, since the portal always answers with the most recent known location.

Based on raw data stream, {\em clean data} stream is produced. This includes deduplicated tuples with denoisified geo-coordinates. 
The key step following it is integrating stream data with timetables to determine whether reported location means that a tram of interest can be considered to be in time, delayed or arriving early. This results in the ultimate data stream containing deduplicated data instances extended with additional features being denoisified coordinates, movement status and delay status, referred to as data stream $U$.

As illustrated in Fig. \ref{fig:streaming}, all the intermediate data streams are persisted in Apache Hadoop HDFS file system. This selection is justified by the volume of daily data of one stream reaching 400-600 MB per a day.
Importantly, it is the $U$ data stream that will be used as an input for stream mining.
This study proposes the machine learning algorithm considered for this processing, to be used in a stream mining module receiving the data via Apache Kafka and producing the predictions into Apache Kafka to make them available for other modules.

\section{Short term delay prediction}
\label{sec:problem}
\begin{figure}
\includegraphics[width=\textwidth]{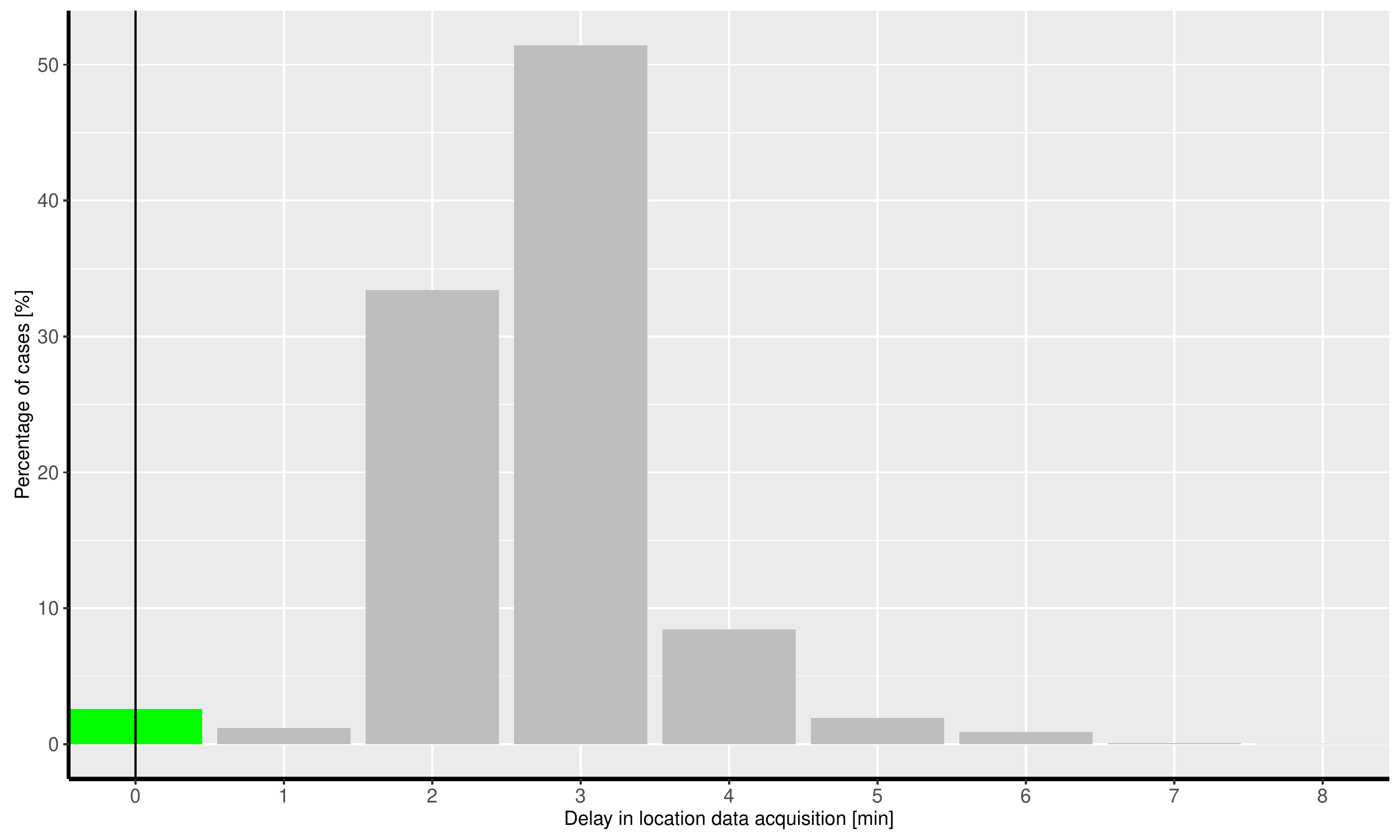}
\caption{The latency of location data collection.}
\label{fig:transmission latency}      
\end{figure}

IoT data including the location data arrives from individual sensors with varied latency. This may be caused by battery conservation needs or obstacles preventing wireless transmission. Moreover, many caching layers may exist between client application and sensor producing the data. In the analysed case, this results in non-negligible latency. 
Fig.~\ref{fig:transmission latency} depicts the proportion of differences $t_\mathrm{R}-t_\mathrm{C}$ observed among location records received for a sample day of 13th of September 2016 and tram location data. It can be observed that most frequently the location data arrives with the latency of 3 minutes. However, communication issues may cause delays up to 8 minutes. 

Such differences raise the need for prediction of current location and dependant data such as delays. What should be noted here is that  location prediction to be done with machine learning techniques, needs extensive data sets covering all vehicle routes.
However, delay prediction is also of major importance e.g. for dynamic connection planning. Delay prediction can be understood as regression problem in which the difference between planned and observed arrival time is calculated. Alternatively, the problem can be treated as a classification task i.e. prediction whether a vehicle of interest will be delayed in $\Delta t$ minutes can be made. 
In both cases the input for the prediction can be the vector of past coordinates of the vehicle $\mathbf{c}^i=[c_{t-K\delta t},\ldots,c_{t-k\delta t}]$, the vector of past delays of vehicle $\mathbf{d}^i=[d_{t-K\delta t},\ldots,d_{t-k\delta t}]$ or the combined vector of both coordinates and delays $[\mathbf{c}^i,\mathbf{d}^i]$.
In the remainder of this study, we will concentrate on the classification case and test $[\mathbf{c}^i,\mathbf{d}^i]$, $\mathbf{c}^i$, and $\mathbf{d}^i$ cases. Moreover, dependent variable will take one of three values \{\texttt{before\_time,delayed,on\_time}\}.

As far as key requirements for short term prediction as concerned, the ability to predict delays based on limited training data is of particular importance. Delay prediction is of key importance when trams are operated at unusual schedules e.g. during large events such as marathons. Hence, the trade-off between the need to collect sufficiently large training data set and the need to perform prediction during the period of training data collection. On the one hand, techniques such as MLP networks are known to be able to model complex, non-linear dependencies. On the other hand, stream mining techniques, where a simplified yet available prediction model can be potentially sufficiently robust after processing just several instances are of potential value.
\section{Hybrid stream mining of delay data streams}
\label{sec:method}
The method proposed in this study relies on a combination of two techniques i.e. multilayer perceptron and stream mining model.
The former of the models will be referred to as $P_N$, where $N$ stands for the number of training instances used to train the model.

In line with \cite{ditzler2015}, let us define data stream as a sequence of instances $\mathcal{S}_1,\mathcal{S}_2,...$ where $\mathcal{S}_k$ denotes  tuple $\mathcal{S}_k=\{(\mathbf{x}_k,\mathit{y}_k)\}$ composed of a feature vector $\mathbf{x_k}\in R^n$ and a discrete class label $y_k$ .
Once, instance $\mathcal{S}_{k}=\{(\mathbf{x}_k,\mathit{y}_k)\}$ arrives in the stream data, loss between predictions $\hat{\mathit{y}_k}$ and true values $\mathit{y}_k$ can be computed. Similarly to most works on stream mining techniques, let us assume that the evaluation of stream mining models will be based on standard {\em prequential} evaluation \cite{bifet2010} i.e. first unlabelled instance is used to get a prediction, next to update a stream mining model.

While stream mining model is constantly evolved based on newly arriving labelled instances $\{(\mathbf{x}_k,\mathit{y}_k)\}$,
the neural model $P_N$ is developed based on the first $N$ instances i.e. the training set $T_N=\{(\mathbf{x}_k,\mathit{y}_k):k=1,\ldots,N\}$.
Before $N$ instances are available, let us propose the prediction of a hybrid model $H_k$ to be fully based on the prediction generated by a stream mining model.
 More precisely, for $\forall k\leq N H_k=h_k$. Moreover, for $k>N$ let us propose the hybrid model to switch between the use of neural prediction and stream model prediction based on the accuracy of the two models observed during sliding window of no more than $L$ preceding instances. In this way, the hybrid model can use the prediction of a model expected to yield higher accuracy, in order to minimise the loss between true and predicted labels.
Based on these properties, let us propose the hybrid algorithm combining stream and neural model, referred to as CSaNN in the remainder of this work and formulated in Alg. \ref{alg:main}. 

\begin{algorithm}
\caption{Hybrid CSaNN classification method
\label{alg:main}}
\KwIn{
$\mathcal{S}_1,\mathcal{S}_2,...,\mathcal{S}_k=\{(\mathbf{x}_k \in R^n,\mathit{y}_k)\},...$ - data stream,
$N$ - the size of training data set to be used to develop neural model, $f$ - feature selection function, $L$ - sliding window size
}
\KwData{
$P_N$ - MLP built with the training data of $N$ instances, having $2n$ neurons in hidden layer, $h_i$ - stream mining model developed after processing $i$ instances, $T$ - training data set for MLP network
}
 \KwResult{
$\mathcal{H}_1,\mathcal{H}_2,\ldots$ - the stream of CSaNN predictions generated for individual instances, $\omega_\mathrm{h}, \omega_\mathrm{P}$ - sliding window-based accuracies of stream and neural model, respectively, $c(k)\in\{0,1\}$ - the category of underlying model used to provide $\mathcal{H}_k$, with 1 standing for stream model, and 0 for neural model, respectively
}
\Begin{
$T=\phi$\;
\For{$i=1,\ldots$}
{
 \uIf{$i\leq N+1$}{
   $T=T\cup f(\mathcal{S}_i)$\;
   $\mathcal{H}_i=h_{i-1}(f(\mathcal{S}_i))$\;
   $h_i$=update$(h_{i-1},f(\mathcal{S}_i))$\;
  
  \If{i=N}{
    $P_N$=train$(T)$ \;
    }
  }
  \Else{
	$\omega_\mathrm{h}=\frac{\sum_{j=max(i-L,1)}^{i-1} (h_{j-1}(f(\mathcal{S}_j))=\mathit{y}_j)}{i-max(i-L,1)}$\;
	$\omega_\mathrm{P}=\frac{\sum_{j=max(i-L,N+1)}^{i-1} (P_N(f(\mathcal{S}_j))=\mathit{y}_j)}{i-max(i-L,N+1)}$\;
	$[\mathcal{H}_i,c(i)] =
  \begin{cases}
   [h_{i-1}(f(\mathcal{S}_i)),1]     & \quad \omega_\mathrm{h}\geq \omega_\mathrm{P}\\
  [P_{N}(f(\mathcal{S}_i)),0] & \quad \text{otherwise}\\
  \end{cases}$
	  }
}
}
\end{algorithm}
An important input for Alg. \ref{alg:main} is feature selection function $f()$, which selects a subset of all instance features to use it both for model development and prediction needs.

\section{Results}
\label{sec:results}

In order to analyse the benefits of the use of MLP networks combined with stream models, two data streams were used i.e. WAW1316 and WAW1923. Both of them contain Warsaw tram data. However, each data stream comes from a different period and includes the data of different tram lines  contained in a varied number of instances, as shown in Table \ref{tab:data_streams}.
Every instance in both streams is developed based on $U$ data stream described in Sect. \ref{sec:fusion}. For every instance $\mathcal{S}_t=\{(\mathbf{x}_t,\mathit{y}_t)\}\in U$, a corresponding instance $\mathcal{F}_t=\{(
\mathbf{x}_{t-K\delta t},\ldots,\mathbf{x}_{t-\delta t},\mathbf{x}_{t},\mathit{y}_{t+\Delta t})\}$ containing the tram data from $K$ preceding time steps is developed. Ground truth class label $\mathit{y}_{t+\Delta t}$
is a delayed label i.e. the label available with the latency of $\Delta t$ minutes. In the study, it is used for the evaluation with the prequential evaluation process documented in Alg. \ref{alg:main}. 
 Importantly instance $\mathcal{F}_t$ is developed out of the data of instances $\mathcal{S}_t=\{(\mathbf{x}_t,\mathit{y}_t)\}\in U$ describing the same tram while it is moving in the same direction. 
Finally, only complete instances are placed in the data streams documented in Table \ref{tab:data_streams}.

\begin{table}
\caption{The data streams used for method evaluation.
\label{tab:data_streams}}
\begin{center}
\begin{tabular}{|c|c|c|r|c|c|c|}
\hline 
Data set & Period & Tram lines & No. of instances & $K$ & $\delta t [sec.]$ & $\Delta t [min.]$\\ 
\hline 
WAW1216 & 12\textsuperscript{th}-16\textsuperscript{th} September 2016 & 9,25 & 10740 & 5& 60 & 5\\ 
\hline 
WAW1923 & 19\textsuperscript{th}-23\textsuperscript{rd} September 2016 & 4,17,18 & 8631 & 5& 60 & 5 \\ 
\hline 
\end{tabular} 
\end{center}
\end{table}

\begin{figure}
\includegraphics[width=\textwidth]{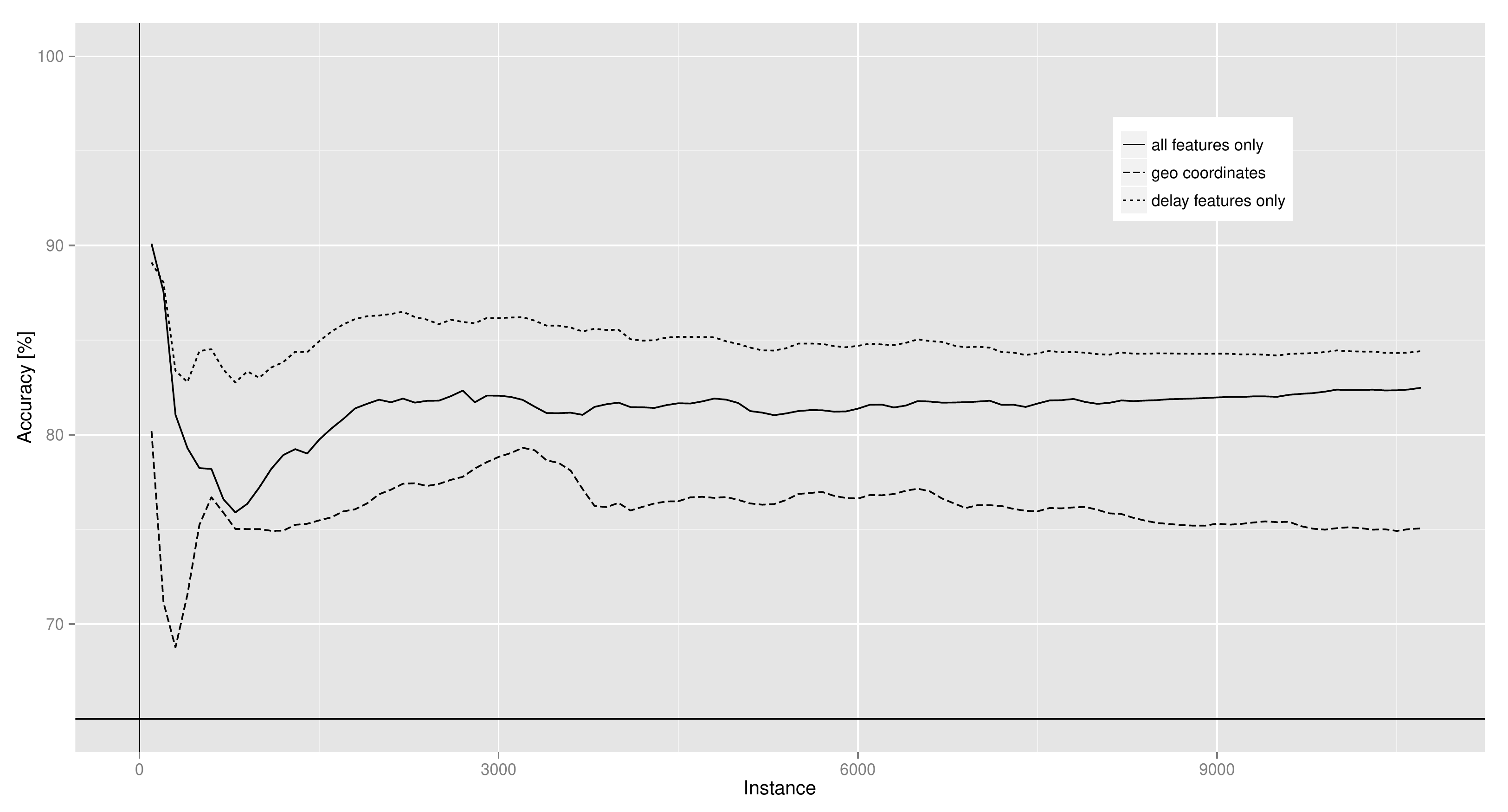}
\caption{The impact of feature selection on the accuracy of streaming models. WAW1216 data stream.}
\label{fig:feature_impact_on_accuracy}      
\end{figure}

In this work, one of the most popular stream classification techniques, namely Hoeffding tree (HT) model  proposed by P. Domingos and H. Hulten in  \cite{domingos2000} is used as a stream mining model.
Before the accuracy of the hybrid method proposed in this study is analysed, let us consider the impact of feature selection on the accuracy of HT.  The result of such analysis is shown in Fig. \ref{fig:feature_impact_on_accuracy}. 
It can be observed that the use of all instance features i.e. both tram coordinates and delay features from $K$ preceding time steps yields worse results than the use of delay features only. The worst results are attained when the input for a model predicting delay status in $\Delta t$ minutes is composed of preceding geo-coordinates of the tram only. Hence, $f=f_\mathrm{D}$ selecting only delay features out of all available features will be used throughout the rest of this work.

\begin{figure}[t]
\includegraphics[width=\textwidth]{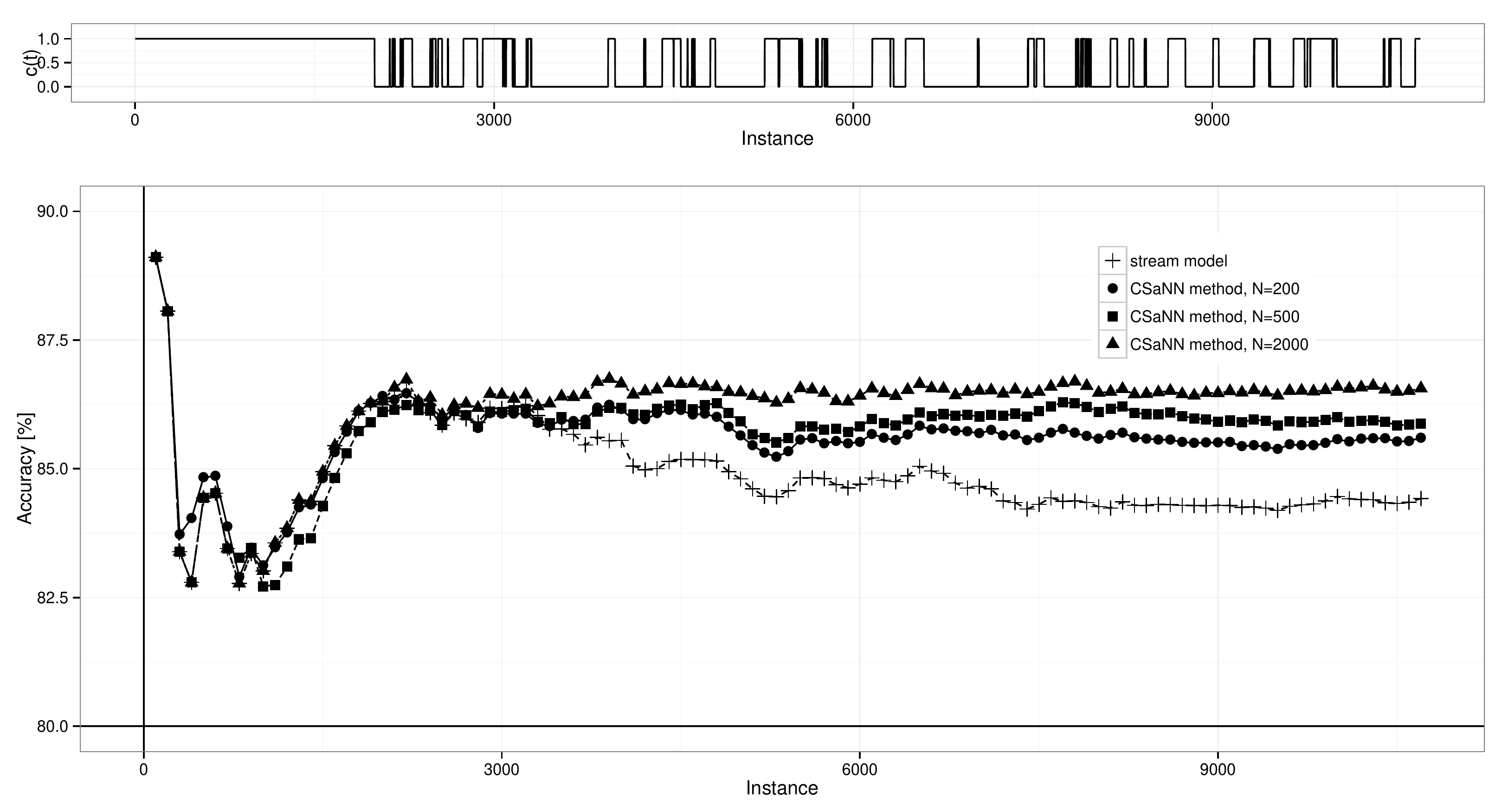}
\caption{Lower part: The comparison of stream model and CSaNN method incorporating MLP network developed with training data sets of different cardinality $N$. Upper part: the source of individual predictions $c(t)$ of CSaNN model ($N$=2000).}
\label{fig:stream_hybrid_all}      
\end{figure}

Let us compare the results of using HT with the accuracy attained with CSaNN method for WAW1216 data stream. The accuracy $A(t)$ is shown in the lower part of Fig. \ref{fig:stream_hybrid_all}. It is calculated for individual prediction techniques as $A(i)=100\times \frac{\sum_{j=1}^{i} (\tilde{\mathit{y}}_j=\mathit{y}_j)}{i}$, where $\tilde{\mathit{y}}$ stands for the prediction generated by the method of interest. The benefits of CSaNN method can be observed for all $N$ settings, with the largest accuracy improvement in the case of $N=2000$ i.e. the largest size of training data set. Importantly, the use of hybrid method proposed in this study prevents accuracy loss, which is observed for stream model i.e. HT. Moreover, larger training data sets yield higher precision predictions. However, the trade off between higher accuracy batch model and the time needed to collect larger training data set has to be considered.

In addition, the upper part of Fig. \ref{fig:stream_hybrid_all} shows for which stream instances, the CSaNN model selected HT-based prediction (denoted with 1), and for which of them the prediction provided by neural model $P_N$ was used as $\mathcal{H}_t$ (which is denoted with 0). It can be observed that the hybrid method proposed in this study frequently switches between the use of stream and batch model based on moving average accuracies $\omega_\mathrm{h}$ and $\omega_\mathrm{P}$, respectively. This clearly confirms that none of the models is superior throughout all the time windows. What is even more important, the overall accuracy plots shown in the lower part of Fig. \ref{fig:stream_hybrid_all} confirm the accuracy gains arising from the hybridisation proposed in this study.


The question arises whether the superiority of CSaNN holds also for another reference data stream. The summary data for both data streams
is provided in Table \ref{tab:comparison}.  It can be observed that in the case of WAW1923 data stream, the improvement of classification accuracy reaches 3.88\% after processing 8000 instances. This is even though the neural model is considered as an alternative only throughout 6000 instances, as $N=2000$ has been applied in the tests.

\begin{table}
\begin{center}
\caption{The comparison of stream model and CSaNN method for different data streams. Results for CSaNN, $N$=2000, $L$=100 are shown.
\label{tab:comparison}}
\begin{tabular}{|c|r|c|c|c|c|c|c|c|}
\hline 
Data set & Method & $A$(1000) & $A$(2000) & $A$(3000) & $A$(4000) & $A$(5000) & $A$(6000) & $A$(8000) \\ 
\hline 
WAW1316 & HT         & 83.02 & 86.31 & 86.17 & 85.55 & 84.80 & 84.70 & 84.26 \\ 
\hline 
• & CSaNN & 83.02 & 86.31 & 86.44 & 86.65 & 86.48 & 86.42 & 86.48 \\ 
\hline 
WAW1923 & HT         & 80.80 & 82.85 & 81.23 & 80.33 & 79.66 & 80.55 & 80.40\\ 
\hline 
• & CSaNN  & 80.80 & 82.85 & 82.90 & 82.48 & 82.60 & 83.47 & 84.28\\ 
\hline 
\end{tabular} 
\end{center}
\end{table}

To sum up, the CSaNN method proposed in this study provides improved prediction accuracy. Importantly, if offers adaptive switching between neural batch model and HT-based stream model. Hence, it reduces the risk of using  batch model not matching recent data stream in the case of concept drift or stream model of lower accuracy than batch model. Therefore, the method can be considered suitable for industrial use.

\section{Conclusions and future work}
\label{sec:conclusions}
Smart objects comprising on IoT deployments frequently report data with latency. Factors such as wireless transmission problems or the need for increased battery lifetime contribute to this problem. Hence, the need for short-term prediction that could serve as a way of imputing missing data instances. In this study, the problem of processing delay data streams for public transport systems is undertaken. A hybrid method used for short term prediction is proposed.

The hybrid model composed of stream and neural models, proposed in this study has been shown to provide increased prediction accuracy. Importantly, the model exhibits the feature of stream mining model at the same time i.e. it can be used for prediction purposes already after processing just a few instances in a data stream. This eliminates the need for extensive training data set required to train neural network, before any predictions can be made.
At the same time, the contribution of soft computing model to a hybrid method proposed in this work is also evident, since the prediction accuracy is increased, which is a trend observed throughout entire period following the training of multilayer perceptron.

In the future, further extensions of the proposed method are planned. In particular,  dynamic feature selection during stream model evolution can be considered. Similarly, multiple neural models based on the training data sets of varied cardinality can be introduced in the hybrid model.

\paragraph{Acknowledgements}
This is a pre-print of a contribution published in In: P\'{e}rez García H., Alfonso-Cend\'{o}n J., S\'{a}nchez Gonz\'{a}lez L., Quinti\'{a}n H., Corchado E. (eds). {\em International Joint Conference SOCO’17-CISIS’17-ICEUTE’17 León, Spain, September 6–-8, 2017, Proceeding. SOCO 2017, CISIS 2017, ICEUTE 2017}. Advances in Intelligent Systems and Computing, vol. 649, published by Springer. The definitive authenticated version is available online via \\ \url{https://doi.org/10.1007/978-3-319-67180-2\_18}.

This research has been supported by the European Union's Horizon 2020
research and innovation programme under grant agreement No. 688380 VaVeL:
Variety, Veracity, VaLue: Handling the Multiplicity of Urban Sensors.


%
%
\bibliographystyle{splncs03}

\begin{thebibliography}{10}
\providecommand{\url}[1]{\texttt{#1}}
\providecommand{\urlprefix}{URL }

\bibitem{almeida2013}
Almeida, E., Ferreira, C., Gama, J.: Adaptive Model Rules from Data Streams,
  pp. 480--492. Springer Berlin Heidelberg, Berlin, Heidelberg (2013)

\bibitem{bifet2010}
Bifet, A., Holmes, G., Kirkby, R., Pfahringer, B.: Moa: Massive online
  analysis. J. Mach. Learn. Res.  11,  1601--1604 (Aug 2010),
  \url{http://dl.acm.org/citation.cfm?id=1756006.1859903}

\bibitem{bifet2009data}
Bifet, A., Kirkby, R.: Data stream mining a practical approach  (2009)

\bibitem{ditzler2015}
Ditzler, G., Roveri, M., Alippi, C., Polikar, R.: Learning in nonstationary
  environments: A survey. IEEE Comp. Int. Mag.  10(4),  12--25 (2015),
  \url{http://dblp.uni-trier.de/db/journals/cim/cim10.html\#DitzlerRAP15}

\bibitem{domingos2000}
Domingos, P., Hulten, G.: Mining high-speed data streams. In: Proceedings of
  the Sixth ACM SIGKDD International Conference on Knowledge Discovery and Data
  Mining. pp. 71--80. KDD '00, ACM, New York, NY, USA (2000),
  \url{http://doi.acm.org/10.1145/347090.347107}

\bibitem{hesse2015}
Hesse, G., Lorenz, M.: Conceptual survey on data stream processing systems. In:
  2015 IEEE 21st International Conference on Parallel and Distributed Systems
  (ICPADS). pp. 797--802 (Dec 2015)

\bibitem{kolozali2014}
Kolozali, S., Bermudez-Edo, M., Puschmann, D., Ganz, F., Barnaghi, P.: A
  knowledge-based approach for real-time iot data stream annotation and
  processing. In: Proceedings of the 2014 IEEE International Conference on
  Internet of Things(iThings), and IEEE Green Computing and Communications
  (GreenCom) and IEEE Cyber, Physical and Social Computing (CPSCom). pp.
  215--222. ITHINGS '14, IEEE Computer Society, Washington, DC, USA (2014),
  \url{http://dx.doi.org/10.1109/iThings.2014.39}

\bibitem{marz2013}
Marz, N.: Big data : principles and best practices of scalable realtime data
  systems. O'Reilly Media, [S.l.] (2013),
  \url{http://www.amazon.de/Big-Data-Principles-Practices-Scalable/dp/1617290343}

\bibitem{qin2016}
Qin, Y., Sheng, Q.Z., Falkner, N.J., Dustdar, S., Wang, H., Vasilakos, A.V.:
  When things matter. J. Netw. Comput. Appl.  64(C),  137--153 (Apr 2016),
  \url{http://dx.doi.org/10.1016/j.jnca.2015.12.016}

\bibitem{tsai2014}
Tsai, C.W., Lai, C.F., Chiang, M.C., Yang, L.T.: Data mining for internet of
  things: A survey. IEEE Communications Surveys Tutorials  16(1),  77--97
  (First 2014)

\end{thebibliography}

\end{document}